# Preseismic electric signals generated by critical stress loading of the lithosphere. A review of the type of signals obtained from a long lasted (1999 to 2012) experiment conducted in Greece.

Thanassoulas[1], C., Klentos[2], V., Verveniotis, G.[3], Zymaris, N.[4]

1. Retired from the Institute for Geology and Mineral Exploration (IGME), Geophysical Department, Athens, Greece.
   e-mail: thandin@otenet.gr - URL: www.earthquakeprediction.gr

2. Athens Water Supply & Sewerage Company (EYDAP),
   e-mail: klenvas@mycosmos.gr - URL: www.earthquakeprediction.gr

3. Retired Ass. Director, Physics Teacher at 2nd Senior High School of Pyrgos, Greece.
   e-mail: gver36@otenet.gr - URL: www.earthquakeprediction.gr

4. Retired, Electronic Engineer.

**Abstract**

In this work a review is made of the various preseismic electric signals that have been observed before large EQs in Greece during the period of 1999 – 2012. The observed preseismic electric signals comply quite well with the theoretical ones that are expected to be generated by a large scale piezoelectric mechanism been activatsd at the focal area due to its excess stress-load.
Preseismic electric signals of the total piezoelectric field, its first time derivative, its oscillation due to M1 and K1 tidal components have been observed by single monitoring sites during the actual 1999 – 2012 experiment in Greece. Moreover, the "strange attractor like" electric preseismic signal was detected by the simultaneously use of two distant monitoring sites. It is demonstrated that the preseismic electric signals can not only determine quite accurately a very short time window for the EQ occurrence and its epicentral location but the latter can be achieved without any prior knowledge of the geological, tectonic setting or past seismic history of the seismogenic area. The predicted pending EQ is treated as a single isolated destructive nature event which sends clear warning signals well before its occurrence. Real examples are presented from the Greek territory.

1. **Introduction.**

In the seismology branch of science the prediction of an earthquake (large and destructive) is considered as its Holly Grail. The term "earthquake prediction" refers to the knowledge of the earthquake prognostic parameters that is the location, the time of occurrence and its magnitude, for some time before it takes place. According to the prognostic time window it is distinguished as: long-term, referring to a time window of some decades of years, medium-term, referring to a time window of a few years (2-3) and short-term, referring to a time window of the order of up to a couple of months, while sometime the term "immediate" is used when the time window is of the order of a few days. EQs have been treated in the past using statistical models (from basics to elegant ones) aiming to the determination of their prognostic descriptive parameters (time of occurrence, location and magnitude).

The earthquake prediction evolution in time is described in details by Geller (1997), who concludes that despite research has been conducted for more than 100 years, "no obvious" success has resulted. The question that arises is what the cause of this failure is. Thanassoulas (2007) referred to this problem in a very general approach and commented it with the use of the following sketch (fig. 1).

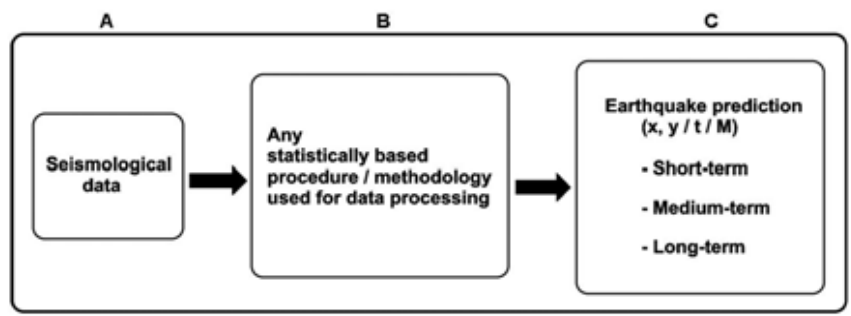

Fig. 1. Generalized flow chart is shown that indicates the procedure which has been followed towards earthquake prediction. A = input data, B = methodology used / physical model etc, C = wanted output, x, y, t, m are the prognostic parameters for all the types of EQ prediction (x, y stand for EQ coordinates in any appropriate coordinating system, t stands for the time of the EQ occurrence and M stands for the EQ magnitude).

Figure (1) shows the generalized procedure, used very often by the various researchers, to statistically solve the earthquake prediction problem. The wanted output (C) that is the prognostic parameters of the output of the system (EQ prediction), depends on the used procedure / methodology (B), and the input data (A). So far this system has presented very limited success or has failed. Therefore, its failure must be attributed either to part (A) or to part (B) or to both of them. It is suggested that, part (B) is highly unlike to fail, since it consists mainly of mathematically validated, robust, statistical methods. At the same time part (A) is valid, too, since it consists of the seismological data, collected, by the different seismological observatories and there is no doubt about their validity. This peculiar non-conformable situation can only be explained with the assumption <u>that part (A) and (B) are not compatible in terms of physical laws</u>. In other words these refer to different physical quantities / procedures that cannot be interrelated with any rational physical / mathematical model.

The incompatibility of figure (1) leads us to the selection: at first of a new data set as (A) and secondly to a different methodology / procedure (B) which will be used for the processing of the input data. The latter, additionally, dictates new physical models to be adopted and to be used in the physical / mathematical analysis of the earthquake prediction problem. If all these are valid and true, then the relation between A, B, C will be a valid one and the problem will have, in principle, been solved. A final remark to be made is that the data used as input must intrinsically, even in a hidden way, convey the information of location, time and magnitude of a future earthquake.

In a very interesting recent paper (Kagan et al. 2012) has been written that: "However, since 1990s numerous statistical tests have failed to support characteristic earthquake and seismic gap/cycle models, and the 2004 Sumatra earthquake and 2011 Tohoku earthquake both ripped through several boundaries between supposed segments" **and moreover** "Earthquake scientists must take a rigorous look at characteristic earthquake models and their implicit assumptions, and should scrap merely long-standing ideas that have been rejected by objective testing or are too vague to be testable".

Consequently, since statistical methods seem to be unable to solve the problem or fail too often, a solution should be sought outside the statistical field of science. Moreover, despite the statistical fail of the Sumatra and Tohoku earthquakes, the latter complied pretty well with the lithospheric plate oscillating model triggered by the M1 tidal component (Thanassoulas et al. 2011, 2011a).

Since the EQs are considered as a rock fracturing phenomenon they follow the main principles of rock mechanics. Therefore, they signify their future occurrence, short before the actual seismic events, by modifying, some time earlier from the actual seismic event, the physical properties status of the corresponding seismogenic area.

Consequently, it seems more efficient to study (in terms of prediction) large EQs <u>as single isolated destructive physical events</u> and further more as been independent from regional geology, tectonics or past seismic history, despite any probable arguments that could be raised and presented by seismologists or geologists.

A physical model that has been adopted, can be initiated and most possibly holds in the seismogenic area, and more specifically in the focal area, is the piezoelectric one. That model was introduced in 80s (Thanassoulas et al. 1986, 1993) while its extensive use was made during a long lasted (1999 – 2012) earthquake prediction experiment conducted in Greece.

In this work we will review the adopted physical model and the corresponding, expected and to be generated precursory electric signals before large EQs in Greece. The observed electric signals during the specific experiment are justified by the adopted physical model and the basic rock mechanics principles.

## 2. The physical working model.

The physical model that has been adopted is the one based on the well-known physical phenomenon of piezoelectricity. Following will be presented its behavior under certain particular stress load conditions of the focal area and the theoretically expected to be generated electric signals. The expected electric signals have to be the result of any proper processing methodology of the "in situ" registered electric raw data (Thanassoulas 2007, Thanassoulas et al. 2009).

### 2.1. Total piezoelectric potential.

The most well - known electric potential generated by piezoelectricity is the one shown in figure (2). The generated potential (b) follows the form of the strain (a) as a function of time. That form of the piezoelectric potential, although it is very easily observed in small scale and small size rock samples in lab experiments, it is quite difficult to observe it directly in nature due to the fact that the sample (focal area) is large and the required technical set-up rather impossible to be physically implemented since it requires very large and extended electrical dipoles installed on ground surface.

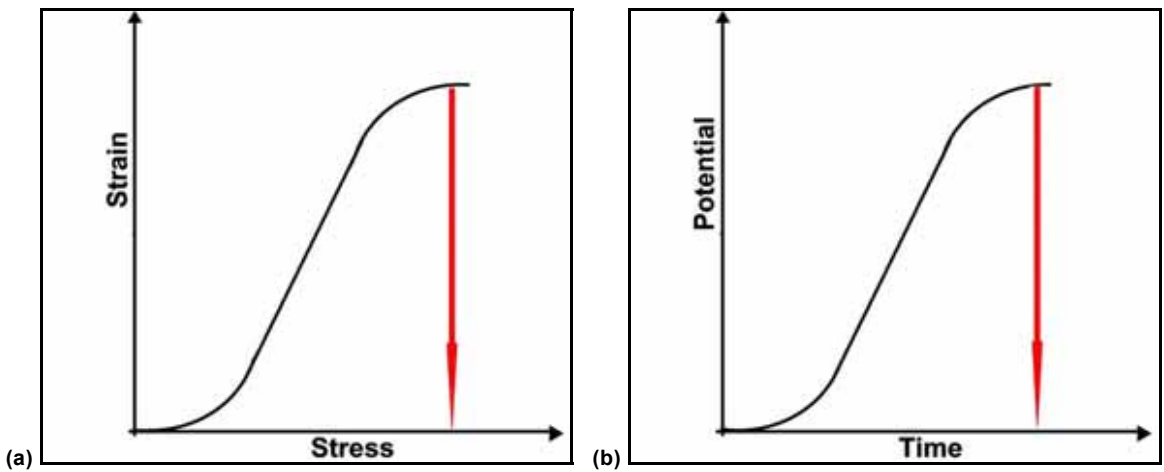

**Fig. 2. Typical stress – strain (a) relation of a solid material is shown that exhibits piezoelectric-like properties and its corresponding potential – time relation (b). It is assumed that positive charges are generated in compression mode. The red vertical arrow indicates the fracturing time of the rock sample.**

Thanassoulas (2007, 2008) presented a solution to that problem. Short electric dipoles were used on ground surface and by integrating along time the registered potential gradient data the potential of figure (2b) was recovered.

If the adopted piezoelectric model is true and valid then electric signals of the form of figure (2b) must be derived from the processing of the registered raw data.

### 2.2. First derivative of the total piezoelectric potential.

Starting from figure (2b) it is clear that the first derivative in time of the piezoelectric potential will have the form of figure (3, lower graph). That type of signals is recoverable from the original raw electric gradient data, after proper integration.

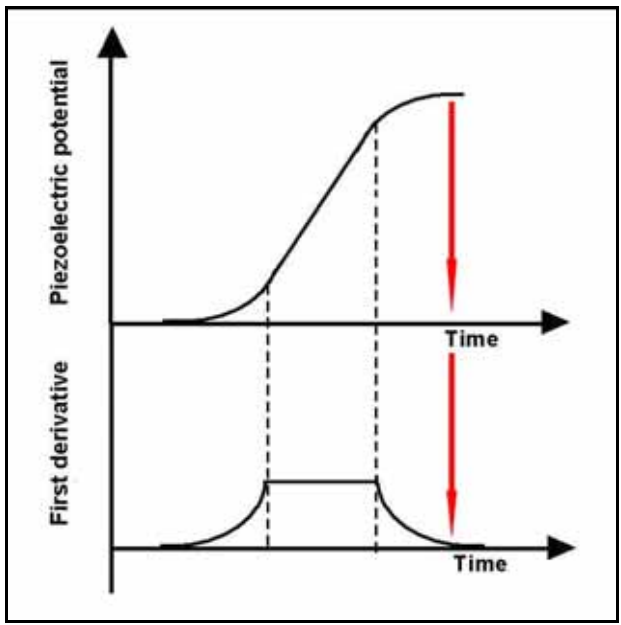

**Fig. 3.** First derivative: Very Long Period (VLP) signals (lower graph) observed through a low-pass filter. The red vertical arrow indicates the fracturing time of the rock sample. The red vertical arrow indicates the fracturing time of the rock sample.

Following figure (3) it is expected that a "plateau" type (Very Long Period, VLP) signal of large duration will be generated well before the occurrence of the future large EQ.

**2.3. Oscillations of the piezoelectric potential of various (tidally triggered) periods.**

Further more, the strain in the focal area presents an oscillating behavior due to the fact that the lithosphere itself oscillates driven by the applied tidal forces upon it (Thanassoulas et al. 1986, 1993). Consequently, the piezoelectric potential will present the same oscillating attitude. The latter is presented in figure (4).

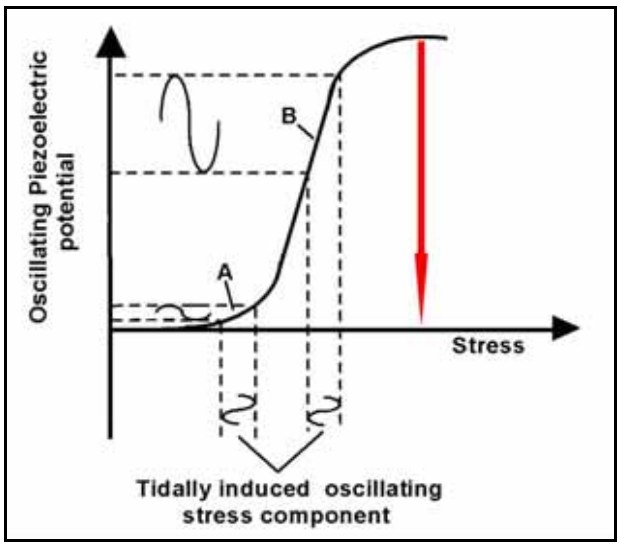

**Fig. 4.** Oscillating signal generated due to oscillating stress component, observed through a band-pass filter (Thanassoulas et al. 1986, 1993). Point A = small strain grad level of the focal area, therefore it results in to a small amplitude oscillating piezoelectric potential component. Point B = large strain grad level of the focal area, therefore it results in a large amplitude oscillating piezoelectric potential component. The red vertical arrow indicates the fracturing time of the rock sample.

Therefore, as long as the focal area is being stress loaded and moves from stress load grad level A to grad level B and approaches the EQ occurrence time, an oscillating electric field must be detected with an increasing amplitude in time.

Actually, the final critical stage of the focal area deformation can be monitored by studying the amplitude increase of the oscillating component of the piezoelectric potential.

**2.4. SES due to small-scale localized pre-fracturing in the focal area.**

Now we consider the two non-linear parts (A, B, fig. 5) of the piezoelectric curve. In area (A) since the rate of change of the stress gradient increases a positive electric pulse is generated. At area (B) the rate of change of the stress gradient decreases and a negative electric pulse is generated. When the time interval between areas (A) and (B) is quite small then a square pulse is generated (Thanassoulas, 2008a). That mechanism is presented in the following figure (5).

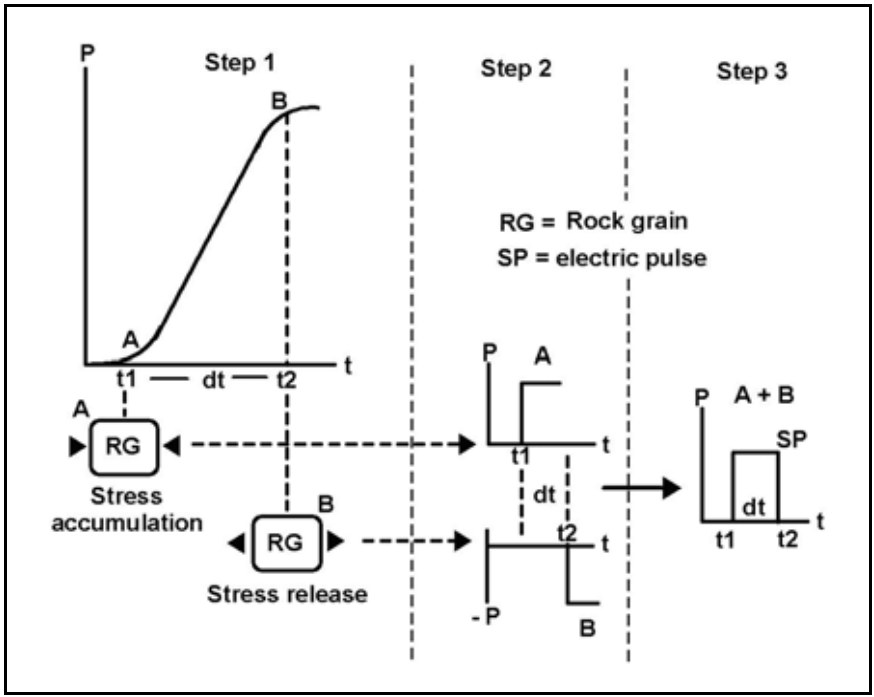

**Fig. 5.** Non-linear areas (A, B) of the piezoelectric potential curve are shown, where piezostimulated (compressional at A and decompressional at B) electric signals can be generated. Step 1, 2, 3: Schematic presentation of the generation of a square electric pulse by the mechanisms of piezoelectricity and piezostimulated currents. A small rock block / grain (RG) is subjected in stress increase at (A) and stress decrease at (B) at step (1). Positive (A) or negative (B) currents generate at step (2) which are combined in step (3) in a short square pulse (SP) of dt duration (after Thanassoulas, 2008a).

That mechanism is activated in the focal area at the early preparatory stage of an earthquake, when micro-fracturing has been initiated and evolves into larger rock fractures that finally form the main fault to be activated during the large seismic event.

Consequently, train - like short electric pulses are expected to be generated at the early stages of the initiation of fracturing of the focal area and short before the main EQ event.

**2.5 The "strange attractor like" electric seismic precursor.**

In paragraph (2.3) it was explained the way the oscillating component of the earth's electric field is generated. In practice, that component is isolated from the raw data by band-pass filtering. Actually, the result of such a filtering procedure is an oscillating electric signal that is composed by any existent same band-pass period electric signal component present in the raw data no matter their origin is. Consequently, the polar diagram of the oscillating electric field intensity vector, at any monitoring site, prescribes an ellipse in the polarization plane (Thanassoulas 2007, 2008b, c, 2009a, b).

In the case of one monitoring site, the oscillating electrical field intensity vector prescribes its specific ellipses, and at any time ($t_i$) it corresponds to an azimuthal direction ($\theta_i$). Therefore, the equation:

$$Y = \theta_i * x + b \qquad (1)$$

is valid for a Cartesian orthogonal system (X, Y) based at the monitoring site where ($\theta_i$) is the angle between the X axis and the electric field intensity vector.

A similar equation, but of different ($\theta_i$), holds for a distant different monitoring site. For the case of two distant different monitoring sites the two linear equations form a linear algebraic system. It is possible to analyze and study, generally, the linear system solutions locations as a function of time. The latter is presented in figure (6) in algebraic form, as follows in the system of equations (2):

$$\left. \begin{matrix} Y_i = X_i * \theta_i(t) + b \\ Y_k = X_k * \theta_k(t) + c \end{matrix} \right|_1^n \blacktriangleright [X, Y]_t \qquad (2)$$

**Fig. 6.** The left part of it represents the system of equations which holds for (n) data pairs obtained from the two (i, k) distant monitoring sites, while the right part of it represents the space of the algebraic system solutions which corresponds the entire electric field registered data pairs.

When the two monitoring sites register random electric oscillating signals then equation (2) generates hyperbolas in the polarization plane. In the case that an earthquake generates an oscillating electric field that can be detected by both monitoring sites then due to the fact that it is detected by both monitoring sites **"in phase"**, an ellipse is generated.

Thus, having adopted that the piezoelectric model is activated in the focal area before large EQs take place, it is expected that observations of the registered earth's electric field prior to large seismic events will result into detecting the type of signals that have been already presented in paragraphs 2.1, 2.2, 2.3, 2.4, and 2.5.

In Greece, during the period of 1999 to 2012, the earth's electric field was continuously registered by monitoring sites, located apart each other at large distances ( www.earthquakeprediction.gr ). The analysis of the recorded raw data revealed that the theoretically expected type of signals had preceded large seismic events. Examples of such signals will be presented as follows.

## 3. Examples of electric signals that preceded large EQs during the 1999-2012 Greek experiment.

### 3.1. Total piezoelectric potential.

**3.1.1. Izmit EQ, August 17$^{th}$, 1999, Mw = 7.6 (VOL monitoring site).**

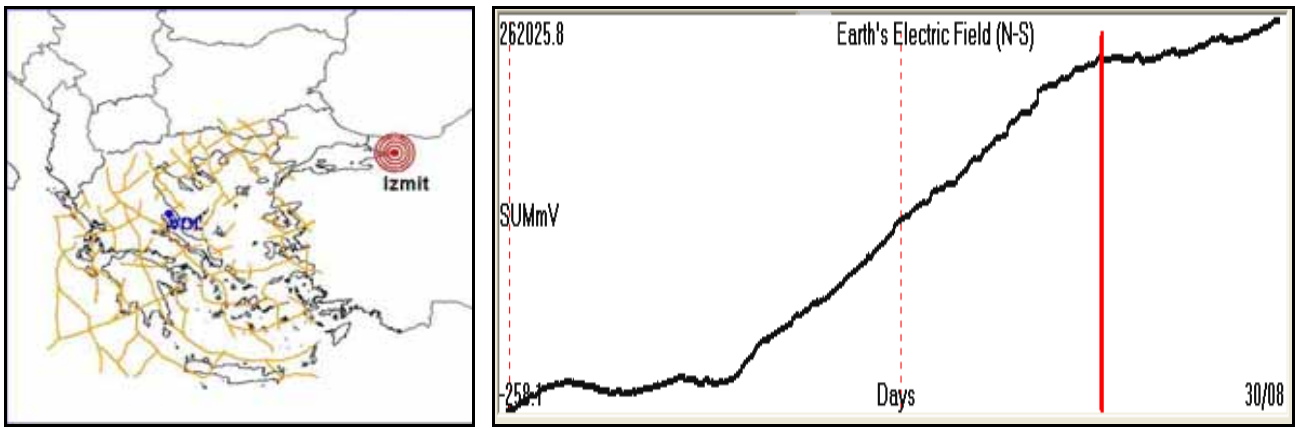

**Fig. 7. Left:** location of **VOL** monitoring site (blue solid circle) and the Izmit EQ (red circles). The thick brown lines represent the deep lithospheric fracture zones and faults of the Greek territory (Thanassoulas, 2007). **Right:** generated potential by the EQ focal area activated piezoelectric mechanism. The Izmit EQ occurrence time is indicated by a vertical red bar.

The form of the determined potential, from the raw grad data registered well before the Izmit EQ, suggests that a large scale piezoelectric mechanism had been activated before that large seismic event.

**3.1.2. Kythira EQ, January 8$^{th}$, 2006, Ms = 6.9R (PYR monitoring site).**

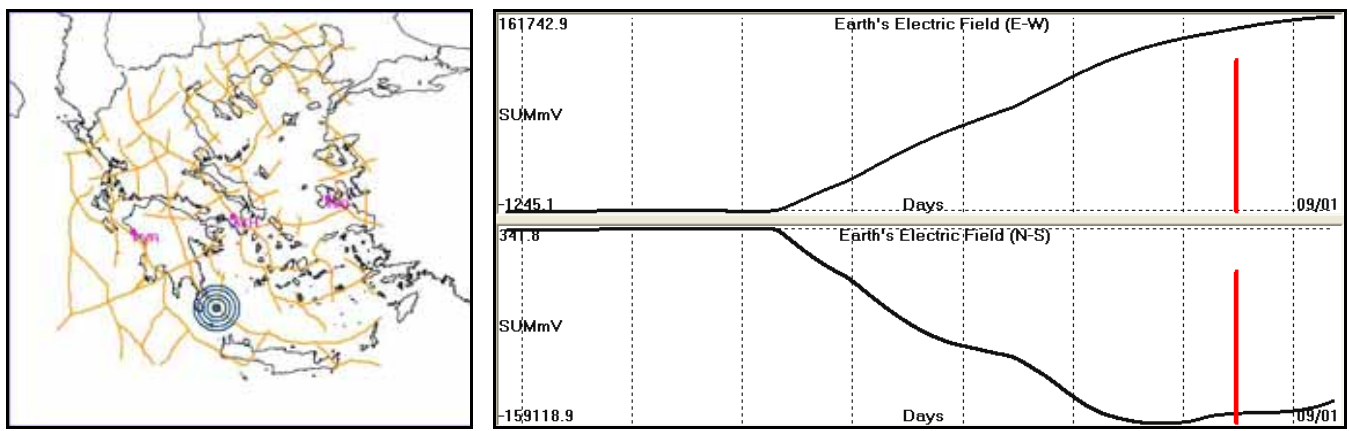

**Fig. 8. Left:** location of the Kythira EQ (blue concentric circles). **Right:** Form of potential (NS and EW components), generated, by the activated piezoelectric mechanism, in the focal area of Kythira EQ. The red bar indicates the time of occurrence of the Kythira, EQ.

In this case too, the form of the determined potential, from the raw grad data registered well before the Kythira EQ, suggests that a large scale piezoelectric mechanism had been activated before that large seismic event.

### 3.1.3. Methoni EQ, February 14th, 2008, Ms = 6.7R (ATH monitoring site).

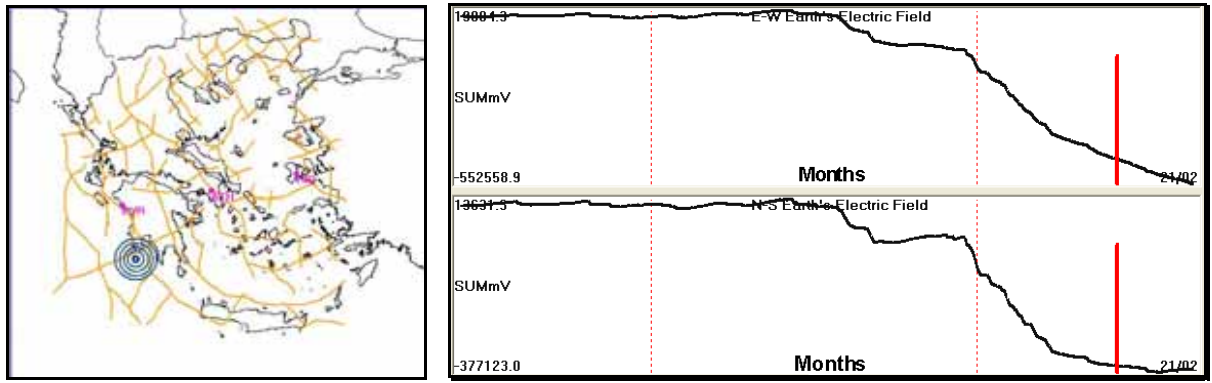

Fig. 9. **Left:** location of the Methoni EQ (blue concentric circles). **Right:** form of potential (NS and EW components), generated, by the activated, physical mechanism, in the focal area of Methoni EQ. The red bar indicates the time of occurrence of the Methoni EQ.

In this case too, the form of the determined potential, from the raw grad data registered well before the Methoni EQ, suggests that a large scale piezoelectric mechanism had been activated before that large seismic event.

### 3.2. First derivative of the total piezoelectric potential.

#### 3.2.1. Izmit EQ. August 17th, 1999, Mw = 7.6 (VOL monitoring site).

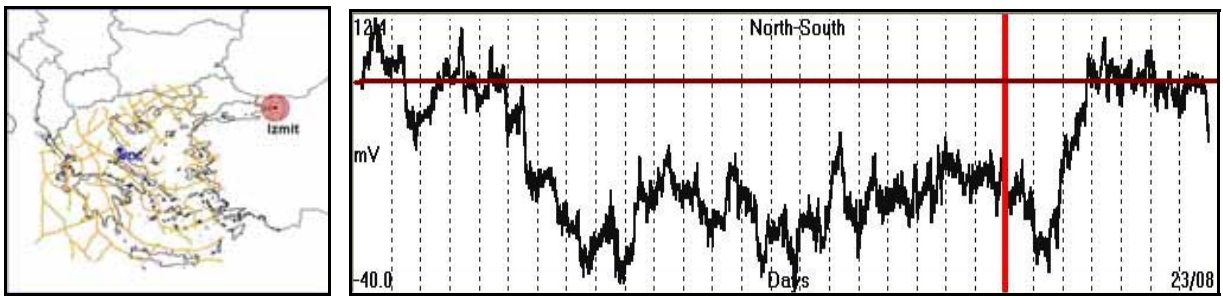

Fig. 10. **Left:** location of VOL monitoring site (blue solid circle) and the Izmit EQ (red circles). The thick brown lines represent the deep lithospheric fracture zones and faults of the Greek territory (Thanassoulas, 2007). **Right:** first derivative electric signal (VLP) determined from the raw gradient recorded data. The brown horizontal line represents the background level of the determined electric field while the vertical red line indicates the EQ occurrence time.

The entire phenomenon was initiated almost 17 days before the EQ occurrence.

#### 3.2.2 Milos EQ, May 21st, 2002, Ms = 5.6R (VOL monitoring site).

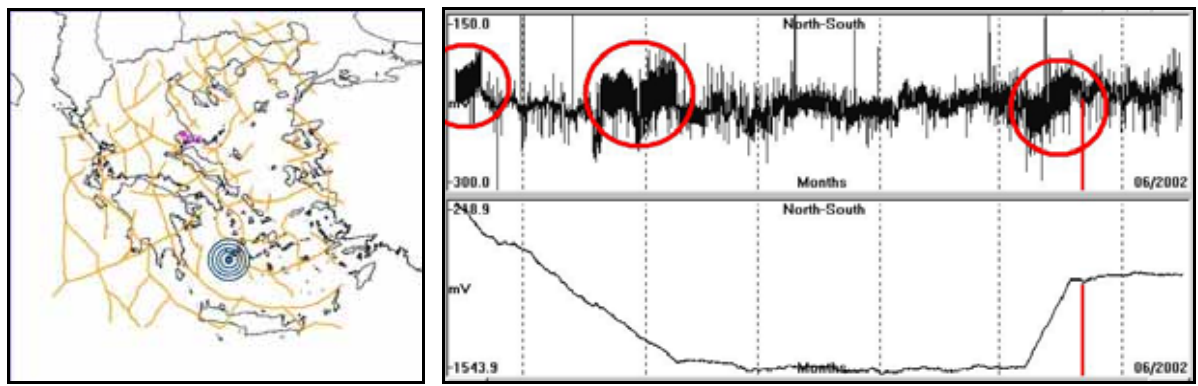

Fig. 11. **Left:** location of Milos EQ is indicated by concentric blue circles. **Lower right:** first derivative electric signal (VLP) determined from the raw gradient recorded data. **Upper right:** registered raw data indicating the presence of SES (in red circles). The EQ occurrence time is indicated by a red vertical line.

The determined large scale piezoelectric potential lasted for almost 4.5 months. SES signals (denoted by red circles), of the paragraph 2.4 type, are observed at the start and end of the corresponding VLP signal.

### 3.2.3. Kythira EQ, January 8th, 2006, Ms = 6.9R (PYR monitoring site)

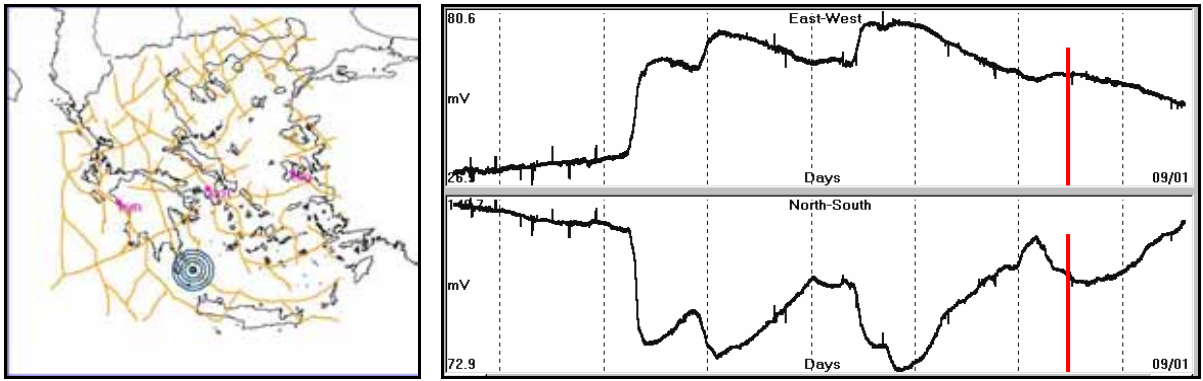

**Fig. 12. Left:** location of Kythira EQ is indicated by concentric blue circles. **Right:** Six (6) days recording of the Earth's electric field, recorded by PYR monitoring before Kythira EQ time of occurrence (red bar).

The VLP signal is visible in both (NS and EW) components of the recorded electric field.

## 3.3. Oscillations of the piezoelectric potential of various (tidally triggered) periods.

### 3.3.1. Skyros EQ. of July 26th, 2001, Ms = 6.1R (VOL monitoring site, K1 tidal component, T = 1 day).

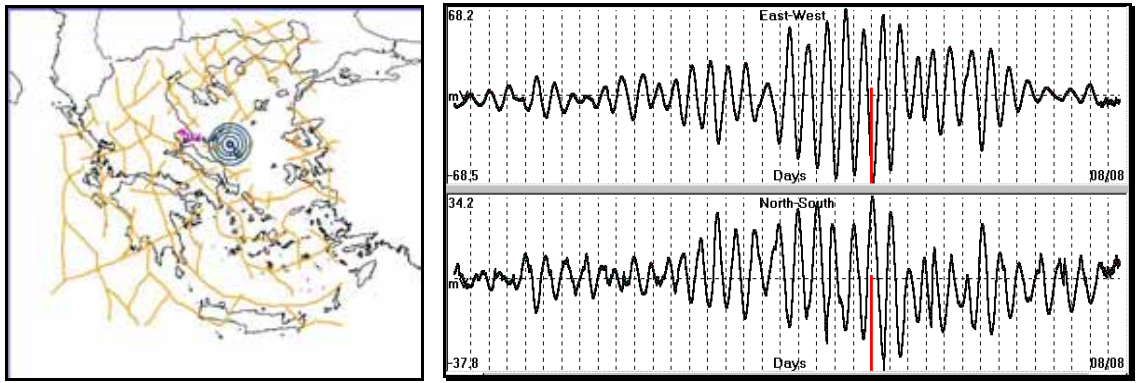

**Fig. 13. Left:** location of Skyros EQ is indicated by concentric blue circles. **Right:** Earth's electric oscillatory field, recorded in VOL monitoring site, prior to Skyros earthquake (Ms = 6.1R) in Greece (2001). The red bar indicates the EQ occurrence time.

The oscillating signal started to evolve almost 11 days before the EQ occurrence.

### 3.3.2. Saros EQ, July 6th, 2003, Ms = 5.9R (PYR monitoring site, K1 tidal component, T = 1 day).

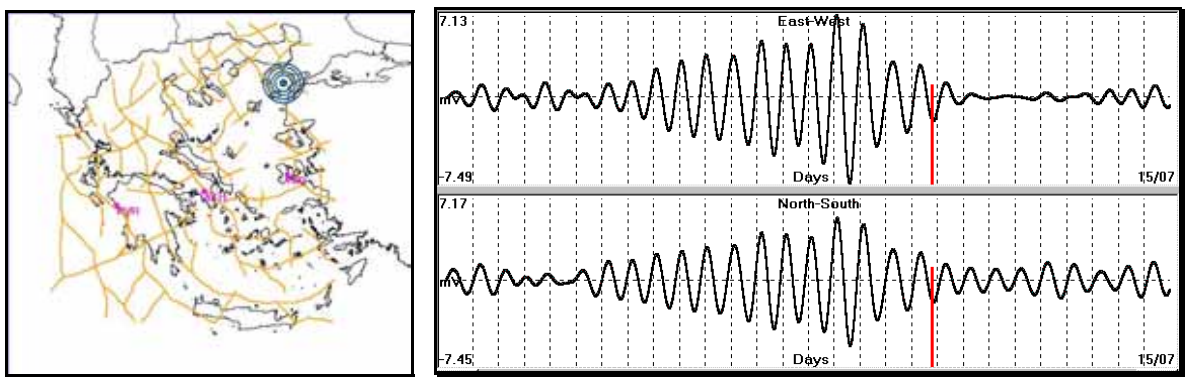

**Fig. 14. Left:** location of Saros EQ is indicated by concentric blue circles. **Right:** Earth's electric oscillatory field, recorded by PYR monitoring site, prior to Saros EQ (Ms = 5.9R) in Western Turkey (2003). The red bar indicates the EQ occurrence time.

The oscillating signal started to emerge 13 days before the EQ occurrence time.

### 3.3.3. Lefkada EQ, August 14th, 2003, Ms = 6.4R (ATH monitoring site, M1 tidal component, T = 14 days).

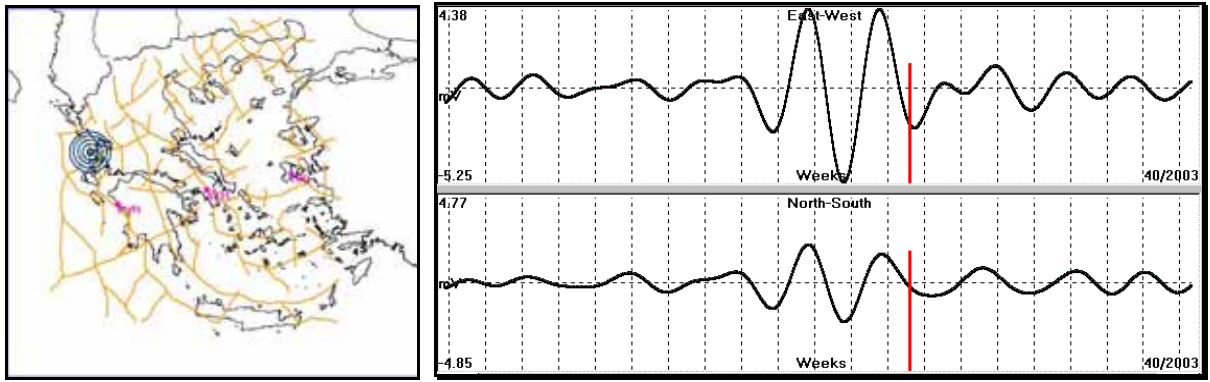

**Fig. 15. Left:** location of Lefkada EQ is indicated by concentric blue circles. **Right:** 14 days period oscillating earth's electric field observed prior to Lefkada, (August 14th, 2003, Ms = 6.4R) earthquake in Greece. A red bar indicates the time of occurrence of this seismic event.

The oscillating signal evolved almost 4 weeks before the EQ occurrence time.

### 3.3.4. Kythira EQ, January 8th, 2006, Ms = 6.9R (PYR monitoring site, M1 tidal component, T = 14 days).

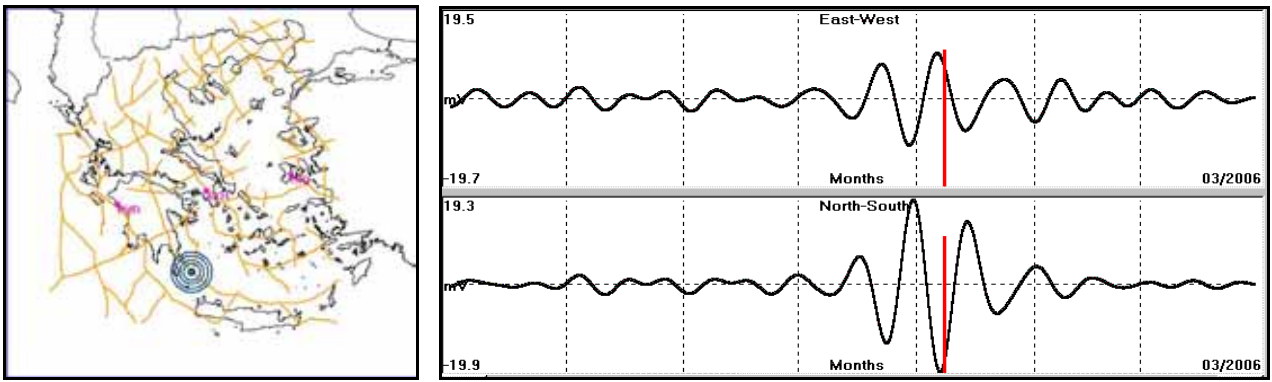

**Fig. 16. Left:** location of Kythira EQ is indicated by concentric blue circles. **Right:** 14 days period oscillating earth's electric field, observed, prior to Kythira, (January 8th, 2006, Ms = 6.9R) earthquake in Greece. A red bar indicates the time of occurrence of this seismic event.

The oscillating signal evolved almost 1 month before the EQ occurrence time.

### 3.4. SES due to small-scale localized pre-fracturing in the focal area.

The following figures (17, 18, 19) show recorded SES at VOL monitoring site before the Izmit EQ occurrence.

**Izmit EQ, August 17th, 1999, Mw = 7.6 (VOL monitoring site).**

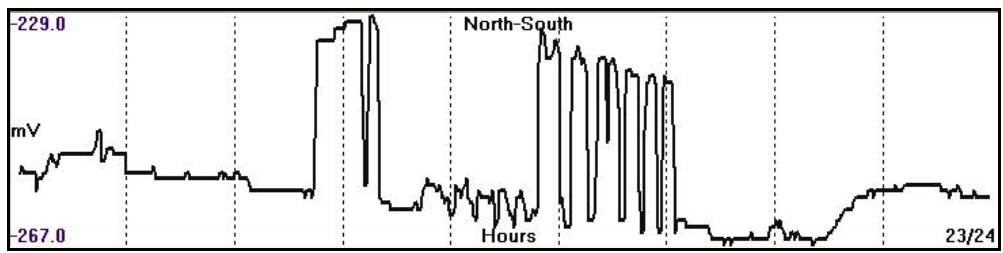

**Fig. 17. SES preseismic electric signal recorded at VOL monitoring site on July 10th, 1999.**

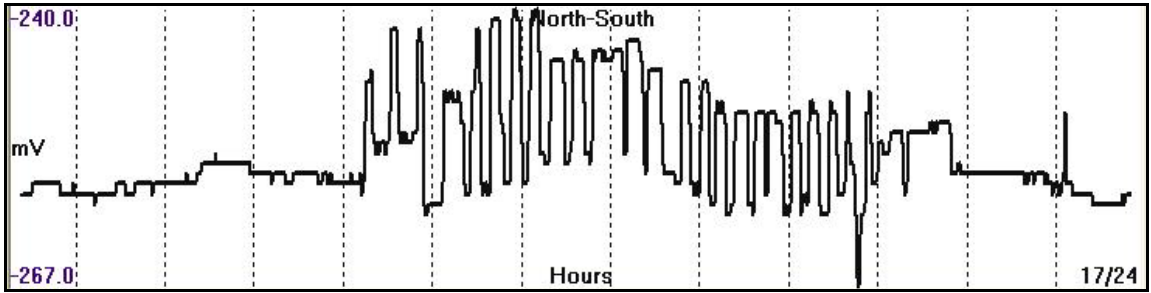

**Fig. 18.** SES preseismic electric signal recorded at **VOL** monitoring site on July 27$^{th}$, 1999.

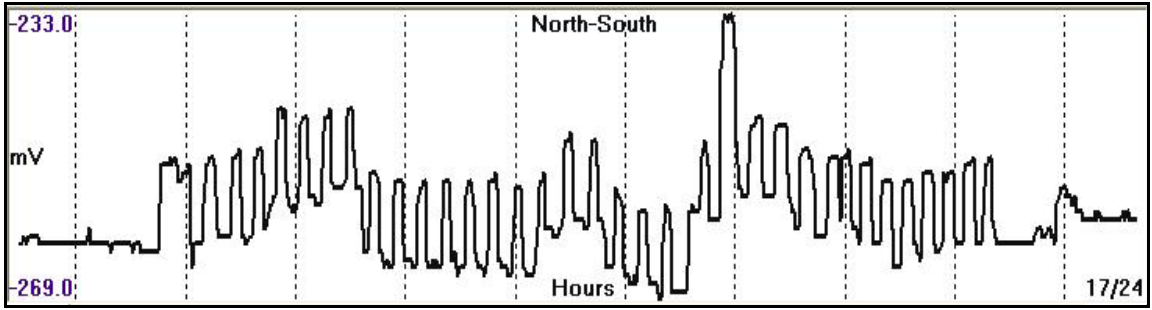

**Fig. 19.** SES preseismic electric signal recorded at **VOL** monitoring site on August 14$^{th}$, 1999.

It is interesting to note from figures 17, 18, 19 that the signal duration increases as long as the EQ occurrence time gets closer (August, 17$^{th}$, 1999).

**Kythira EQ, January 8$^{th}$, 2006, Ms = 6.9R (ATH monitoring site)**

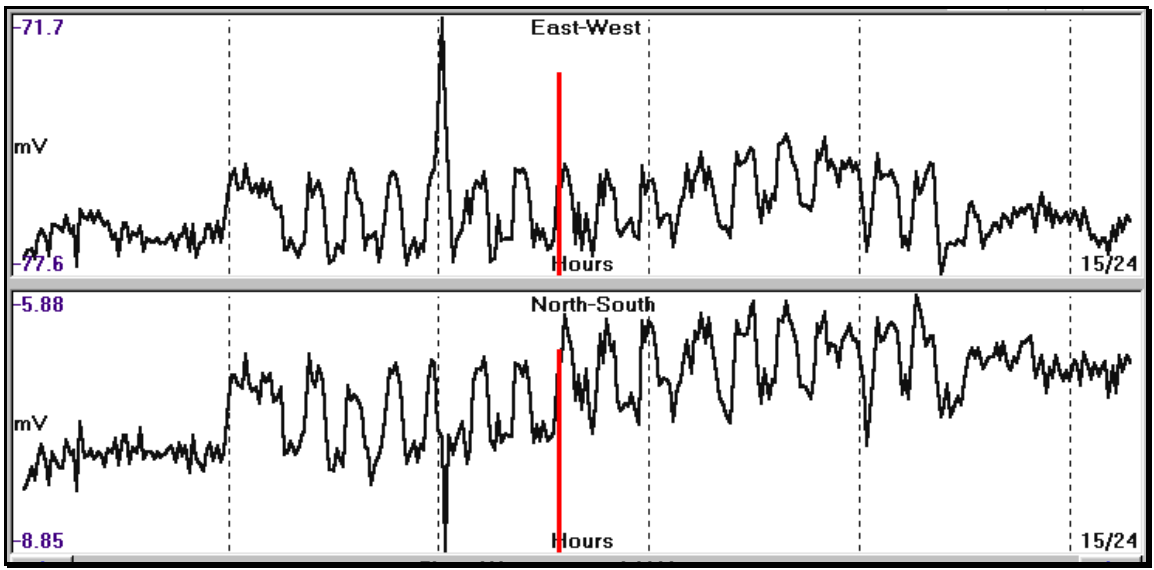

**Fig. 20.** SES precursory, electrical signal recorded, by **ATH** monitoring site, on January 8$^{th}$, 2006. The red bar indicates the time of occurrence of the Ms = 6.9R EQ.

The SES shown by figure (20) is actually a partly preseismic and partly postseismic (concurrent?) signal that started almost 1.5 hours before the EQ occurrence time and ended about 1.5 hours after the seismic event. It presumably lasted all along the "intense fracturing period of time" of the main EQ event.
Finally, figure (21) presents a typical SES recorded by the HIO monitoring site.

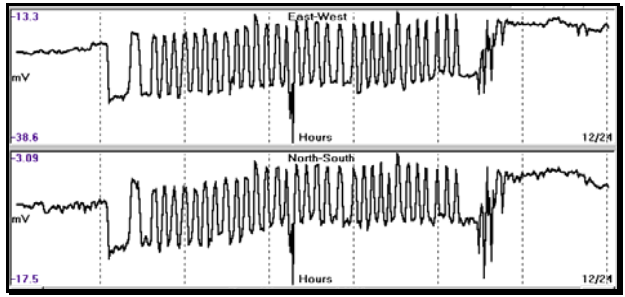

**Fig. 21.** SES precursory, electrical signal recorded, by HIO monitoring site, on May 13th, 2006.

**3.5. "Strange attractor like" preseismic electric precursors.**

Apart from the individual preseismic electric signals presented in paragraphs 3.1, 3.2, 3.3, 3.4, examples are shown from the interaction of the same oscillating preseismic electric field (strange attractor like precursor) at two different and distant monitoring sites as follows.

**Earthquake on April 19th, 2007, Ms = 5.4R.**

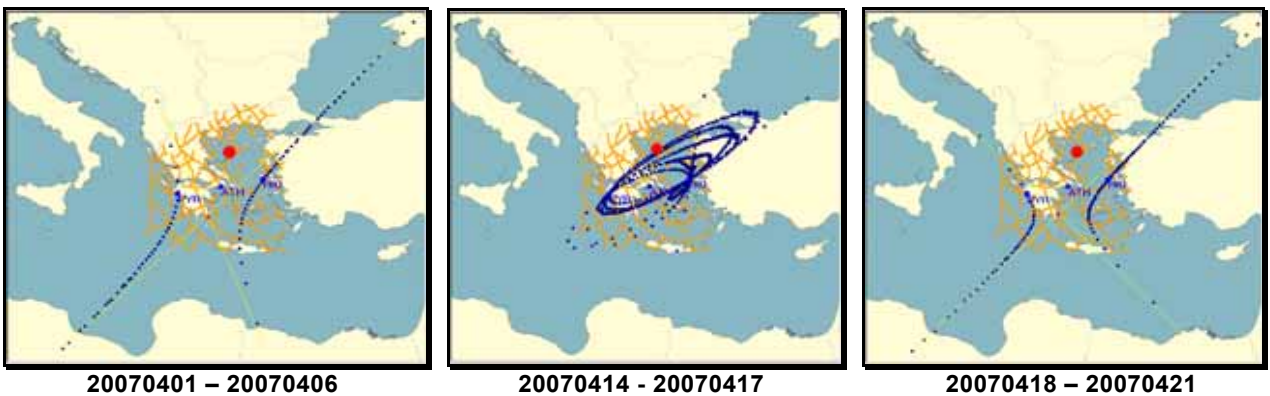

**20070401 – 20070406**    **20070414 - 20070417**    **20070418 – 20070421**

**Fig. 22. Left:** Typical intensity vector intersections are presented valid for period from 20070401 to 20070406. The large, solid, red circle indicates the location of the earthquake (Ms = 5.4R), that occurred on April 19th, 2007. The hyperbolas suggest the absent interrelationship of the recorded signals (no preseismic electric signal is present). **Middle:** recording period = 20070414 – 20070417. The presence of ellipses indicates that the seismogenic area has reached critical stress-load conditions just before the EQ occurrence. **Right:** recording period = 20070418 – 20070421. It is pointed out that the "preseismic signal effect" has vanished 1 day before the EQ occurrence.

The sequence: Hyperbolas – Ellipses – Hyperbolas characterizes the generating cycle of the preseismic "strange attractor like" earthquake precursor that preceded the earthquake (Ms = 5.4R) that occurred on April 19th, 2007.

**Earthquake on February 14th, 2008, Ms = 6.7R.**

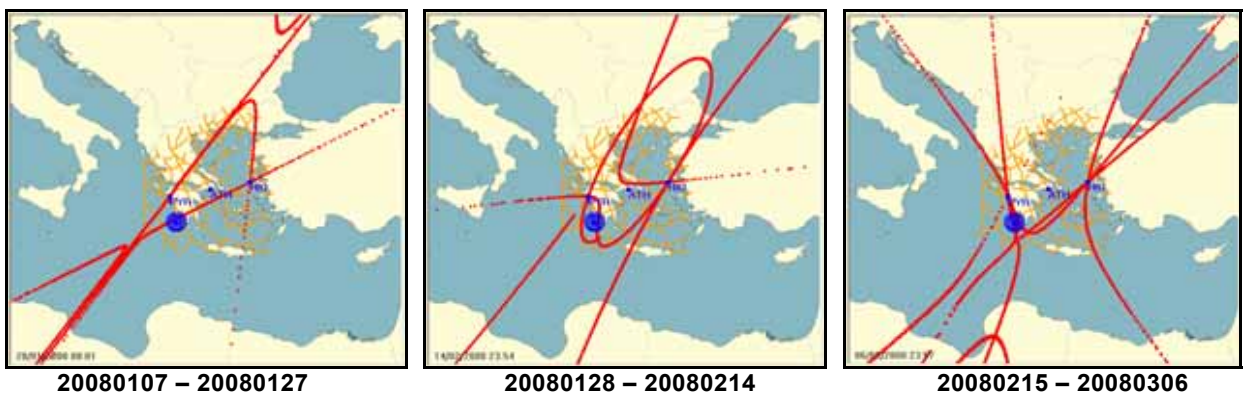

**20080107 – 20080127**    **20080128 – 20080214**    **20080215 – 20080306**

**Fig. 23. Left:** recording period = 20080107 – 20080127. Hyperbolas indicate the absence of the "strange attractor like" preseismic signal. **Middle:** recording period = 20080128 – 20080214. The presence of ellipses indicates that the seismogenic area has reached critical stress-load conditions just before the EQ occurrence. **Right:** recording period = 20080215 – 20080306. The preseismic signal has vanished after the EQ occurrence. The blue concentric circles denote the EQ location.

In this case too, the sequence: Hyperbolas – Ellipses – Hyperbolas characterizes the generating cycle of the preseismic "strange attractor like" earthquake precursor that preceded the EQ of February 14th, 2008, Ms = 6.7R.

**Earthquake on October 14th, 2008, Ms = 6.1R.**

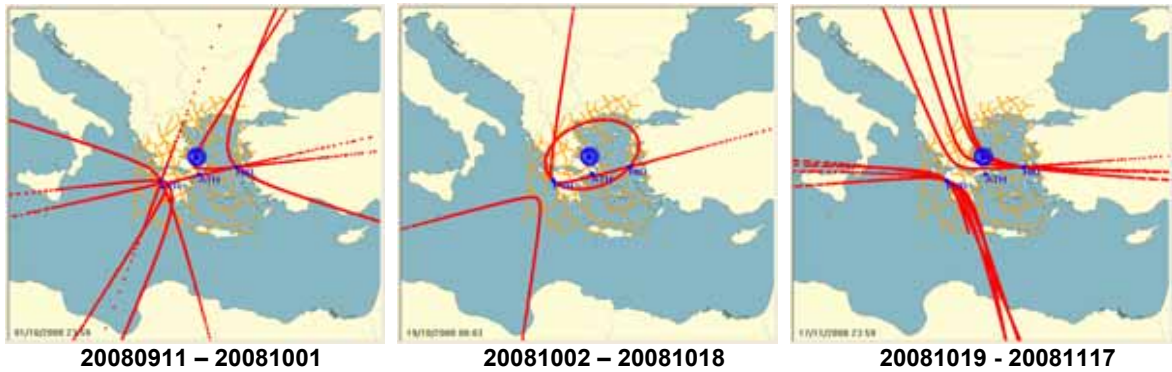

**20080911 – 20081001**  **20081002 – 20081018**  **20081019 - 20081117**

**Fig. 24. Left:** recording period = 20080911 – 20081001. Hyperbolas indicate the absence of the "strange attractor like" preseismic signal. **Middle:** recording period = 20081002 – 20081018. The presence of ellipses indicates that the seismogenic area has reached critical stress-load conditions just before the EQ occurrence. **Right:** recording period = 20080215 – 20080306. The preseismic signal has vanished after the EQ occurrence. The blue concentric circles denote the EQ location.

In this case too, the sequence: Hyperbolas – Ellipses – Hyperbolas characterizes the generating cycle of the preseismic "strange attractor like" earthquake precursor that preceded the EQ of October 14th, 2008, Ms = 6.1R.

**Earthquake on June 8th, 2008, Ms = 7.0R.**

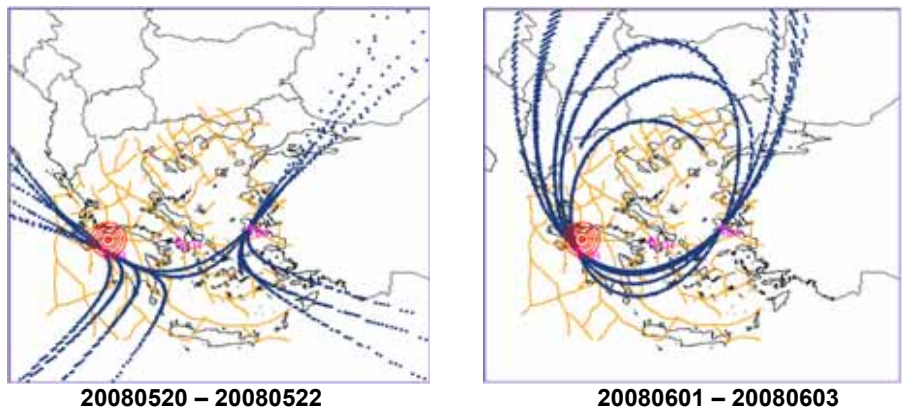

**20080520 – 20080522**  **20080601 – 20080603**

**Fig. 25. Left:** recording period = 20080520 – 20080522. The seismogenic area has not reached critical stress-load conditions since hyperbolas are only present. **Right:** recording period = 20080601 – 20080603. The seismogenic area has entered critical stress load conditions since ellipses have developed.

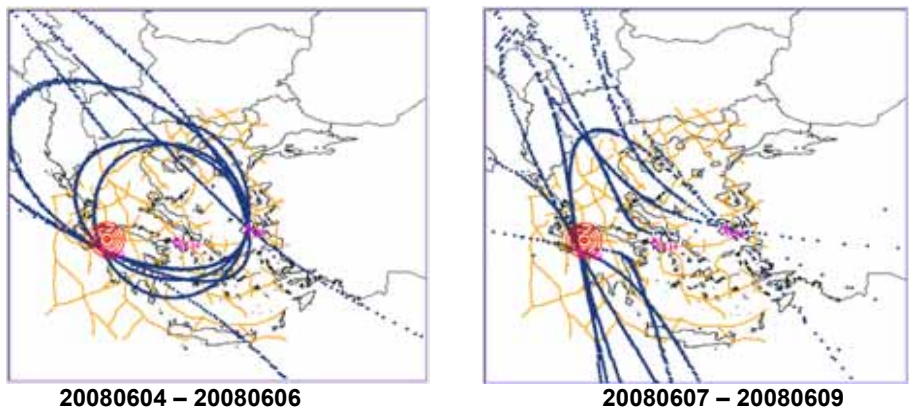

**20080604 – 20080606**  **20080607 – 20080609**

**Fig. 26. Left:** recording period = 20080604 – 20080606. The seismogenic area is still under critical stress-load conditions since ellipses are present. **Right:** recording period = 20080607 – 20080609. . Hyperbolas are present. The preseismic signal has vanished after the EQ occurrence. The red concentric circles denote the EQ location.

In this case too, the sequence: Hyperbolas – Ellipses – Hyperbolas characterizes the generating cycle of the preseismic "strange attractor like" earthquake precursor that preceded the Earthquake on June 8<sup>th</sup>, 2008, Ms = 7.0R. Earthquake precursor vanished 1 day before the main seismic event.

4. **Conclusions.**

A general research methodology followed in all branches of science is the following. In order to study a physical phenomenon a theoretical model is postulated. Then the theoretically generated observations are compared to the real ones observed in nature. Upon agreement of the theoretical to the real observations the postulated model is accepted as a valid one. In the opposite case the postulated model must be corrected or rejected. Along the same line Kagan et al. (2012) referred to as "A precept of science is that theories unsupported by observations and experiments must be corrected or rejected, however intuitively appealing they might be." while Jordan (2006) pointed out that "a scientifically valid hypothesis must be prospectively testable". Generally, the latter methodology has been followed in seismology.

In this work the piezoelectric mechanism has been accepted as the one been activated in the focal area under critical stress load conditions that are met a short time before the occurrence of a large EQ. The analysis of the adopted model justifies the generation of various piezoelectric signals of characteristic forms. Those specific forms of signals have been observed prior to large earthquakes in the Greek territory. Therefore, since the observations of the EQ precursory signals, in nature, comply with the theoretical ones, the adopted physical mechanism must be considered as a valid one.

Furthermore, the analysis of the observed EQ precursory electric field, since it follows basic laws of the electric potential field theories, should provide the location of its generation mechanism by simple triangulation, provided that it has been registered by two at least different and distant monitoring sites. The latter is demonstrated by the following figures (27, 28).

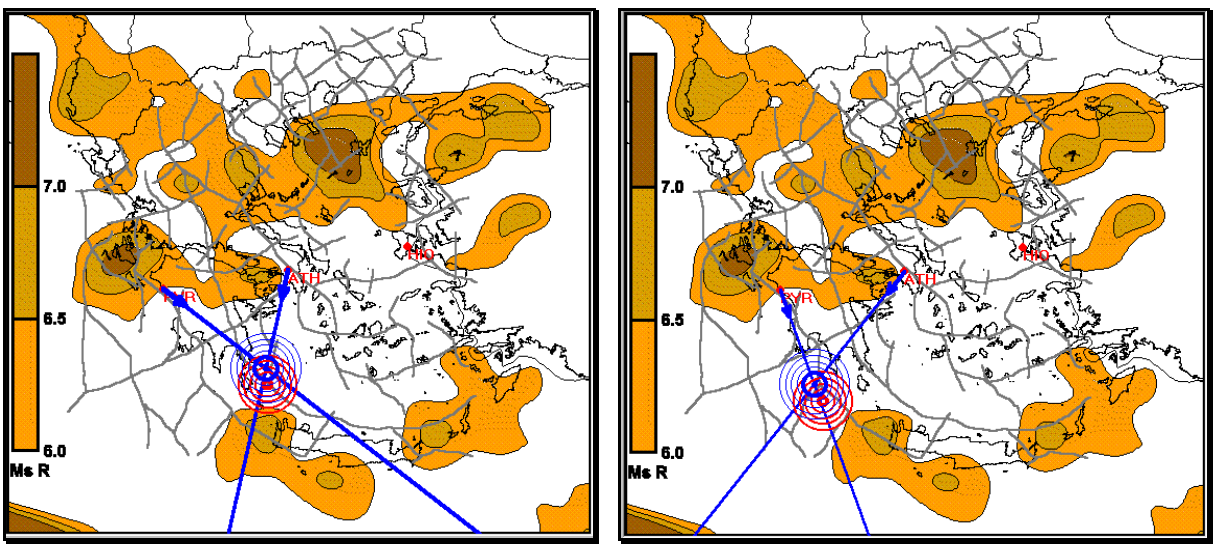

**Fig. 27.** Calculated EQ epicentre (blue circles) in relation to the seismological (red circles) one. **Left = 20060108 EQ, Ms = 6.9R EQ. Right = 20031017 EQ, Ms = 5.8R.** Blue lines = electric field intensity vector calculated at each monitoring site.

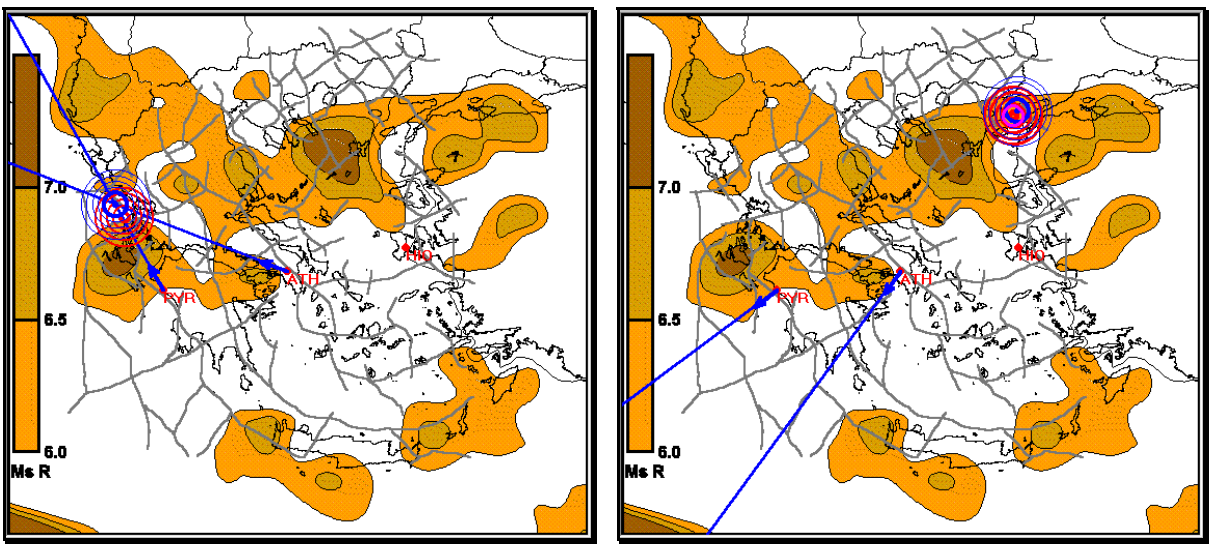

**Fig. 28.** Calculated, epicenter (blue circles) in relation to the seismological (red circles) one. **Left = 20030814 EQ, Ms = 6.3R. Right = 20030706 EQ, Ms = 5.9R.** Blue lines = electric field intensity vector calculated at each monitoring site.

In conclusion, it has been demonstrated that the earthquake electric precursory signals are a very powerful tool for the implementation of a successful earthquake prediction. It can narrow the time window of the occurrence of a large EQ not only to the required one for what is referred to as "short-tem prediction" but even shorter down to an hour at some cases (Thanassoulas et al. 2010). Moreover, it can provide the determination of the location of the future pending EQ quite accurately. It could be said that the study of the EQ precursor electric signals can provide simultaneously two out of three of the predicting parameters of an earthquake, namely the time of its occurrence and its location. The third parameter, the magnitude, can be provided by another physical model (LSEFM) the lithospheric seismic energy flow model (Thanassoulas 2007).

Retrospection of some large EQs (Parkfield, Northridge, Sumatra and Tohoku) suggests that Parkfield, and Northridge EQs conform very well as far as it concerns their magnitude to the postulated lithospheric seismic energy flow model (LSEFM, Thanassoulas, 2007) while Sumatra and Tohoku comply very well, as far as it concerns the day of their occurrence (Thanassoulas et al. 2011a), to the oscillating lithospheric model triggered by the M1 tidal waves. It is important to point out that the presented model is quite efficient in terms of prediction without the need of any geological, tectonic setting or past seismic history prerequisites.

Finally it seems that deterministic methods based on plain physics and physical models can provide more precise earthquake prediction solutions than the statistical ones. We believe that much more effort should be put by the scientific community towards that research direction.

For those who are interested in that specific topic, a detailed description of the total experiment (hardware set-up, registered data, methods for processing the registered data, earthquake catalog and specific examples) are available free for down load from the URL: www.earthquakeprediction.gr. Some more detailed work regarding the topic of the earthquake prediction can be found in www.arXiv.org

## 5. References.

Geller, J, R., 1997. Earthquake prediction: a critical review, Geophys. J. Int., 131, 425-450.

Jordan, T. H. 2006. Earthquake predictability, brick by brick. Seismological Research Letters 77(1), 3-6.

Kagan, Y., Jackson, D., Geller, R. 2012. Characteristic earthquake model, 1884 -- 2011, R.I.P, arXiv.org / 1207.4836v1 [physics.geo-ph].

Thanassoulas, C., 2007. Short-term Earthquake Prediction, H. Dounias & Co, Athens, Greece. ISBN No: 978-960-930268-5.

Thanassoulas, C. 2008. "Short-term time prediction" of large EQs by the use of "Large Scale Piezoelectricity" generated by the focal areas loaded with excess stress load. arXiv:0806.0360v1 [physics.geo-ph].

Thanassoulas, C. 2008a. Pre-Seismic Electrical Signals (SES) generation and their relation to the lithospheric tidal oscillations K2, S2, M1 (T = 12hours / 14 days). arXiv:0806.2465v1 [physics.geo-ph].

Thanassoulas, C., Klentos, V., Verveniotis, G., 2008b. On a preseismic electric field "strange attractor" like precursor observed short before large earthquakes. arXiv:0810.0242v1 [physics.geo-ph].

Thanassoulas, C., Klentos, V., Verveniotis, G., Zymaris, N. 2008c. Preseismic electric field "strange attractor" like precursor analysis applied on large (Ms > 5.5R) EQs, which occurred in Greece during December 1st, 2007 - April 30th, 2008. arXiv:0810.4784v1 [physics.geo-ph].

Thanassoulas, C., Klentos, V., Verveniotis, G., Zymaris, N. 2009. Seismic electric precursors observed prior to the 6.3R EQ of July 1st 2009, Greece and their use in short-term earthquake prediction. arXiv:0908.4186v1 [physics.geo-ph].

Thanassoulas, C., Klentos, V., Verveniotis, G., Zymaris, N. 2009a. Preseismic oscillating electric field "strange attractor" like precursor, of T=14 days, triggered by M1 tidal wave. Application on large (Ms > 6.0R) EQs in Greece (March 18th, 2006 - November 17th, 2008). arXiv:0901.0467v1 [physics.geo-ph].

Thanassoulas, C., Klentos, V., Verveniotis, G., Zymaris, N. 2009b. Preseismic oscillating electric field "strange attractor like" precursor, of T = 6 months, triggered by Ssa tidal wave. Application on large (Ms > 6.0R) EQs in Greece (October 1st, 2006 - December 2nd, 2008). arXiv:0901.4285v1 [physics.geo-ph].

Thanassoulas, C., Klentos, V. 2010. How "Short" a "Short-term earthquake prediction" can be? A review of the case of Skyros Island,Greece, EQ (26/7/2001, Ms = 6.1 R). arXiv:1002.2162v1 [physics.geo-ph].

Thanassoulas, C., Klentos, V., Verveniotis, G., Zymaris, N. 2011. The Japan earthquake of March 11th, 2011 (Mw = 8.9R) as viewed in terms of local lithospheric oscillation due to M1 and K1 tidal components. A brief presentation. arXiv: 1103.2385v1 [physics.geo-ph].

Thanassoulas, C., Klentos, V., Verveniotis, G., Zymaris, N. 2011a. Can large (M >= 8) EQs be triggered by tidal (M1) waves? An analysis of the global seismicity that occurred during 1901 – 2011. arXiv:1106.1081v1 [physics.geo-ph].

Thanassoulas, C., Tselentis, G.A., 1986. Observed periodic variations of the Earth electric field prior to two earthquakes in N. Greece., In: Proc. 8th European Conf. Earthquake Engineering, Lisbon, Portugal, Vol. 1, pp.41-48.

Thanassoulas, C., Tselentis, G., 1993. Periodic variations in the earth's electric field as earthquake precursors: results from recent experiments in Greece. Tectonophysics, 224, 103-111.

URL: www.earthquakeprediction.gr.